\documentstyle[prb,aps,multicol,epsf]{revtex}
\input psfig
\def\be{\begin{equation}}
\def\ee{\end{equation}}
\def\half{\textstyle{1\over 2}}
\def\D{\Delta}

\begin{document}
\bibliographystyle{simpl1}

\title{Entropy and Time}

\author{Vinay Ambegaokar and Aashish Clerk}
\address{Laboratory of Atomic and Solid State Physics,
Cornell University, Ithaca, New York 14853}
\maketitle
\bigskip

\begin{abstract}
The emergence of a direction of time in statistical mechanics from
an underlying time-reversal-invariant dynamics is explained by
examining a simple model. The manner in which time-reversal
symmetry is preserved and the role of initial conditions are
emphasized. An extension of the model to finite temperatures is also
discussed.
\end{abstract}

\section{Introduction}

The second law of thermodynamics is usually stated as the inequality
$\D S\ge 0$ for an isolated system, where $S$ is the entropy. 
Implicit in thermodynamics is thus a direction of time determined
by the evolution toward equilibrium of a macroscopic system with no
external influences.  The extent to
which this notion emerges from statistical mechanics based on an
underlying time-reversal-invariant dynamics is the topic of this
paper.

The point of view taken here is not controversial; it has
been accepted since the work of Boltzmann\cite{BO} was understood. 
Including the topic in this special issue may thus seem unnecessary.
However,
our impression is that many undergraduate (and even many graduate)
courses do not cover this thought provoking topic adequately. The
reason may be that the message can be lost in
subtleties described by words such as {\it Stosszahlansatz,
Umkehreinwand}, and {\it Wiederkehreinwand}, and by discussing the
topic in the context of Boltzmann's equation, Liouville's theorem,
and coarse graining. An alternative is to discuss simple concrete
models. We will consider one that is easy to simulate
and has a long history: the 1907 double-urn model of Paul and
Tatiana Ehrenfest.\cite{PTE} Our treatment is based in part on
Chapter 10 of Ref.~\ref{bib:VA}.

The outline of the paper is as follows. After a brief general
discussion, the way in which a direction of time follows from the
statistics of large numbers is illustrated using the Ehrenfest
model (in its dog-flea version). The difference between time
symmetric fluctuations and the time evolution of a macroscopically
identifiable non-equilibrium initial condition is emphasized.
Calculations are done for a single system and for an
ensemble of systems, the latter being described by a Markovian
equation. With the passage of time (no pun intended), the
model has become more topical than its conceivers could have
imagined. Here it is used to describe the approach to equilibrium of
two-level quantum systems such as spins.
Temperature is introduced into the model via a
Metropolis algorithm, and the approach to equilibrium at constant
temperature is discussed, including a population-inversion (negative
temperature) initial condition. To our knowledge, the Ehrenfest
model has not been used in this way, especially as regards the
introduction of temperature.

\section{Background}

From the point of view of thermal physics, the state of an isolated
physical system with many degrees of freedom is specified by its
energy and other macroscopic parameters such as volume and
magnetization. The assumption of many degrees of freedom implies a
dense spectrum of excitations, and thus a very large number of
microscopic states in a small energy interval consistent
with the given macroscopic parameters. The starting point of a
statistical analysis of a mechanical system is the enumeration
(allowed in principle by both quantum and classical mechanics) of
these microscopic states. The fundamental postulate of equilibrium
statistical mechanics is that each of them is equally probable in
equilibrium; the logarithm of their number is the entropy
associated with the thermodynamic equilibrium of the isolated
system.

We are interested in how equilibrium is reached. A reasonable, but
incorrect, expectation is that if the microscopic states are not
equally likely at some instant, the evolution will be toward a
situation in which they are. If the number of microstates
explored by the system increased with time, its logarithm or
entropy would also increase, giving a statistical underpinning to
the rule that an increase of entropy characterizes spontaneous
processes in isolated systems.

The trouble with this too simple idea is that not every one of the
microscopic configurations consistent with a given macroscopic
non-equilibrium state tends, under the action of the laws of
mechanics, toward equilibrium, although most of them do. As a
result the strict inequality in the second law
has to be replaced by a statement of overwhelming likelihood in
statistical mechanics, thereby allowing the latter to be
consistent with the time-symmetric equations of motion. A tiny
loophole now opens, with the consequence that there is no longer
a strict logical connection between the direction of time and the
increase of entropy: one can never rule
out the overwhelmingly unlikely possibility that a low entropy
initial condition is a time-symmetric giant fluctuation caught
midway. To close the loophole, we have to require that macroscopic
deviations from equilibrium are due to externally imposed initial
conditions.

The word ``overwhelming'' is not being used lightly. To illustrate
its meaning, consider the ratio of the number of microscopic
configurations for a gas filling all or 99.99\% of a container. If
we treat the $N$ molecules of the gas as very weakly interacting, an
estimate for this ratio is $0.9999^{-N}$, because each molecule has
0.01\% fewer available states in the smaller volume.  For a liter,
$N$ is order $10^{22}$, so that the logarithm of the ratio is
$10^{18}$. The reciprocal of the ratio is thus very small indeed.
Yet, this unimaginably tiny number is the probability that a gas in
equilibrium in the entire container would be found to be occupying
99.99\% of its volume, and thus to have undergone a small but
macroscopic entropy-reducing spontaneous fluctuation.

\section{THE DOG-FLEA MODEL}

Treating the evolution of a reasonably realistic statistical system
is technically difficult. Even for a weakly interacting gas, 
collisions cannot be ignored, because they are the mechanism that
shuffles a particular molecule between its states of motion. The
approach to equilibrium thus depends on the details of the motion,
and is a less general phenomenon than equilibrium itself. In order
not to lose the woods for the trees, it
is useful to look for simple but illustrative examples. One such
example is a collection of ``two-level''systems, that is, systems
described by two possible outcomes. 

As a physical realization of this model, one could consider a
collection of weakly interacting quantum spin $\half$ systems each
of which has up and down states. A more whimsical illustration is based
on the model proposed in Ref.~\ref{bib:PTE}: consider a
system consisting of a subsystem of 50 fleas whose ``states''
are residence on
dog A or dog B, which we shall call Anik and Burnside,
sleeping side by side. To simulate molecular agitation, we
suppose that the fleas jump back and forth between the dogs. Now we
need something that plays the role of the rest of the system. Let
us suppose that the fleas are each equipped with
a number, and have been trained to jump when their number is called. The
``environment'' agitating the fleas, which is like a heat
reservoir, will be something that calls out numbers at random,
and our closed system will be the fleas and the reservoir. 

To simulate the approach to equilibrium, it is necessary to start
from a configuration that almost never occurs in the maximally
random state of affairs. Suppose that in the beginning Anik
has no fleas at all. We agitate the fleas by having
a computer generate random numbers between 1 and 50 and
transferring the flea with this number from one dog to the
other. At every step we record the total number (between 0 and 50)
of fleas on Anik, but not their labels. This way has the practical
advantage that we do not have to keep track of the $2^{50}$ ways of
assigning 50 labelled fleas to the two dogs. It also means that we
are following only the ``macrostate.'' 

\begin{figure}
\centerline{\psfig{figure=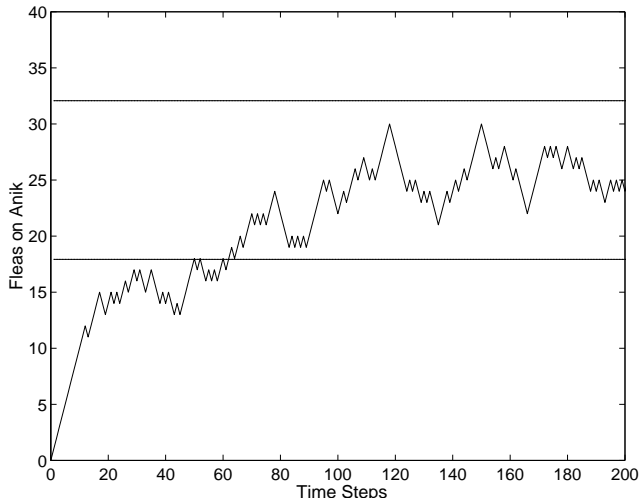,width=8.5cm}}
\caption{
Early time development of the number of fleas on Anik due to random
jumps.  The horizontal lines indicate two standard deviations
above and below the mean.}
\end{figure}

A typical sequence is shown in Figs.~1 and 2. In the first step it
is certain that the
number called will belong to a flea on Burnside. In the second step
the probability of this happening again is 49/50. Thus,
there initially seems to be a steady march towards an equal
partition of the fleas between the dogs. The early time development
is shown in Fig.~1. After about 50 steps
we reach a situation where sometimes Anik and sometimes Burnside
has more fleas. In this region we would expect that every one of
the $2^{50}$ configurations mentioned above would be equally likely.
This expectation translates into a binomial distribution
corresponding to the repeatable random event of tossing 50
fair coins, namely
\be
\label{0}
P(m)={1\over 2^{50}}{50\choose m},
\ee
where $P(m)$ is the probability of the macrostate with $m$ fleas on Anik, and
${50\choose m}$ is the combinatorial coefficient. 

\begin{figure}
\centerline{\psfig{figure=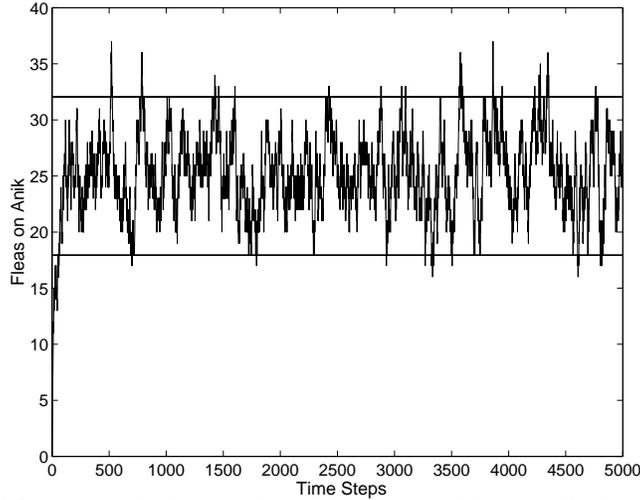,width=8.5cm} }
\caption{Long time behavior of fleas on Anik showing fluctuations in
equilibrium.  The horizontal lines indicate two standard deviations
above and below the mean.}
\end{figure}

An examination of Fig.~2 confirms this expectation. 
The horizontal lines have been drawn to include two standard
deviations on either side of the mean, which corresponds approximately to
the $95 \%$
range for the distribution Eq.~(\ref{0}).
(The standard deviation is $\sqrt{50}/2 \approx
3.5$.) Our eye tells us that, except for the initial transient,
fluctuations outside this range are indeed rare. 

We can be more quantitative. Fig.~3 is a histogram of the relative
durations of the possible outcomes, constructed from Fig.~2 with
the first 100 steps omitted. Superimposed on the histogram is the
binomial distribution. The agreement is very good.\cite{FTNT} The
dance of the fleas in Fig.~2 has thus very quickly forgotten its
unusual starting point and become the endless jitterbug of
``equilibrium,'' in which an event as unlikely as a flea-less dog
simply never happens again without outside intervention.

\begin{figure}
\centerline{\psfig{figure=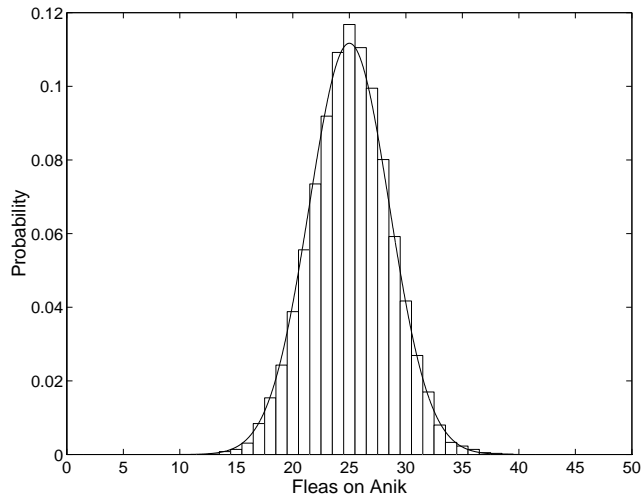,width=8.5cm} }
\caption{
Histogram generated from Fig.~2 with the first hundred time steps
omitted.}
\end{figure}

Figure 2 also illustrates the role of motion reversal.
The model is time reversal invariant because a
string of random numbers in reverse order is just as random. After
the first 100 or so steps, the evolution has no sense of time. 
If we were to expand the region near one of the reasonably large
fluctuations away from the mean, we would find that there is no
characteristic feature of the build-up preceding the maximum
deviation to distinguish it from the time reverse of the decay 
following the maximum. There is also no conclusive argument to
rule out the possibility that the start of the trace shown in Fig.~1
has captured a truly giant fluctuation midway. Of course, we
know that  the figure was not produced in this way. 

The point of the previous paragraph is worth reemphasizing. 
The time asymmetry expressed in the second law is {\it not}
simply the consequence of applying time-symmetric microscopic dynamics
to systems having many  degrees of freedom. Time-reversal symmetry
is indeed preserved in our system. If at some instant the system is in a 
highly improbable state (that is, no fleas on Anik), it is
overwhelmingly likely that it will be in a more evenly-distributed
state at some later time. However, {\it if} the improbable state
were due to a giant fluctuation, precisely the same argument could
be made regarding the prior history of the system -- 
it would then be overwhelmingly likely that at earlier
times the system was also in a more evenly-distributed state. In
this sense, there is perfect symmetry between past and future.

The notion of a statistical ``arrow of time'' thus depends  on the
added ingredient of imposed initial conditions.  When we see a system 
in a highly unlikely state,  we
justifiably assume that this state is the result of  a prepared
starting condition, and not of an overwhelmingly improbable
fluctuation from  equilibrium. As has been particularly emphasized
by Peierls,\cite {PEIERLS}  this setting of initial conditions at
some specified time breaks the symmetry between past and future.

In fact it is virtually impossible to wait long enough for the initial
configuration in Figs.~1 and 2 to occur as a fluctuation in
equilibrium, where it has a probability of $2^{-50}$. To have a
reasonable chance of witnessing such a fluctuation, we would have
to allow a number of time steps approximately equal to the
reciprocal of this probability --- about $10^{15}$ --- to
elapse. Thus, to recover the unlikely configuration of a totally clean
Anik by random shuffling of fleas between equally dirty dogs, 
even for this {\it very} small system of 50 fleas, we would need
a plot roughly two hundred thousand million times as long as
Fig.~2, which extends only for 5000 time steps. Because Fig.~2 is
about 5 cm wide, the length of the required trace would be about
ten million kilometers. In comparison, the distance to the moon is
only about four hundred thousand kilometers. The law of large
numbers is at work, here making an unlikely event overwhelmingly
unlikely. Though not logically certain, it is roughly
99.999999999999999\% probable that the time in Fig.~1 is running in
the direction of increasing disorder.

Ehrenfests' dogs bring into focus the essential characteristics of
time in statistical mechanics. (i) A starting point
macroscopically distinct from equilibrium
is overwhelmingly likely to evolve to greater disorder, that is,
towards equilibrium. (ii) In equilibrium, fluctuations have no
sense of time. (iii) Giant fluctuations from equilibrium to
extremely unlikely states are extremely rare. (iv) A statistically
determined direction of time follows from
the assumption that a system in a highly unlikely ordered 
state has been so prepared by external influences.
Even for the rather small system we considered, 
these uses of the words ``overwhelmingly'' and ``extremely'' are
very conservative.

\begin{figure}
\centerline{\psfig{figure=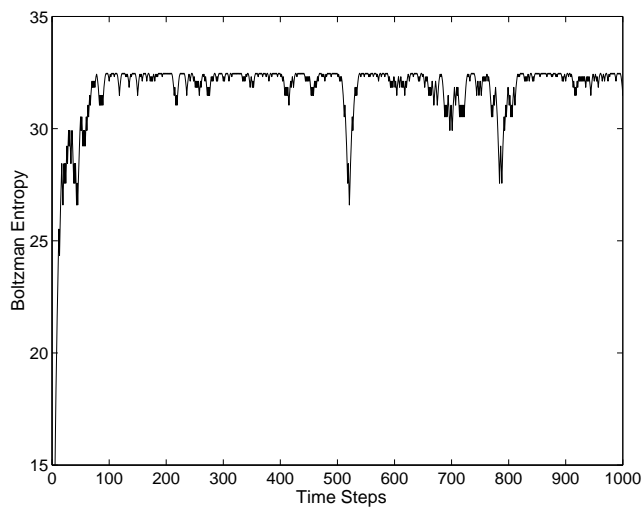,width=8.5cm} }
\caption{
The Boltzmann entropy associated with Fig.~2, showing fluctuations
in equilibrium.}
\end{figure}

The word entropy has not appeared in this section. As a
matter of fact, there is more than one way to introduce that
notion  here, as will be seen in more detail in Section~IV.
The essential point can be made by noting that the states with $n$
fleas on Anik are ``macrostates,'' each allowing for
${50\choose n}$ assignments of distinct fleas or ``microstates.''
We could simply call the logarithm of the latter number, the
entropy, that is,
$S(n)=\ln {50\choose n}$. The combinatorial coefficients have a
maximum half way, at $n=25$,  and become smaller in either
direction. In Fig.~4, the values of $n$ plotted in Fig.~2 have been
converted to a plot of entropy versus time steps using this
rule. As in Fig.~2, $S$ starts from zero, because $S(0)=\ln
{50\choose 0} = \ln 1 = 0$, and has fluctuations. 

This entropy, sometimes called the Boltzmann entropy, can be associated with
a {\it single} time trace such as in Fig.~2.  Although it
fluctuates in equilibrium, the fluctuations diminish as the size of
the system increases.  In a sufficiently large system the
Boltzmann entropy increases steadily as equilibrium is approached. 

Note that we have ignored any contribution to the entropy of the
closed system from the reservoir which is responsible for the 
hopping of the fleas. This assumption is justified here because
energy has not entered into our considerations, making the model
slightly artificial. It is probably best to think of the reservoir
as having a very high temperature (in energy units) compared to the
characteristic energies of the subsystem. As a result, heat
exchanges with the subsystem occur with no change in the entropy of
the reservoir, and only the flea entropy changes with time. The
introduction of energy and temperature in Section~V will lead to an
interesting difference.

\section{GIBBS ENTROPY}

The dog-flea  model is simple enough to allow the solution of
several other interesting problems in time-dependent statistical
mechanics. We first reexamine the assignment of entropy to our
subsystem of fleas. The usual expression for the entropy in
statistical mechanics is
\be
\label{1}
 S = - \sum_i P_i\ln P_i,
\ee
where the sum is over microstates labeled by the index $i$ and
$P_i$ is the probability of $i$. This entropy is associated with a
distribution describing an ensemble of systems, whereas the
Boltzmann entropy introduced earlier is defined for the macroscopic
time development of a single system. If there are
$M$ equally likely microstates, each of the $P$s would equal
$1/M$, and Eq.~(\ref{1}) reduces to $\ln M$. The Boltzmann
entropy has exactly this form if the macrostate with $n$ fleas has
${50\choose n}$ equally likely microstates. As we saw, the
equilibrium Boltzmann entropy fluctuates for the 
subsystem plus reservoir. 

It is possible to assign a constant entropy to equilibrium. A system
of 50 two-level systems at a temperature much higher than the level
spacing is commonly assigned an entropy of $50\ln 2 = 34.657$.
The expression (\ref{1}) gives this result if each of the $2^{50}$
microstates is taken to be equally probable. As we
saw from Fig.~3, the fluctuations in Figs.~1 and 2 are an
expression of equal likelihood of all these microstates. 

Even when the probabilities $P(m)$ of the macrostates corresponding to $m$ fleas
on Anik are not given by Eq.~(\ref{0}), the probabilities of the
equally likely
${50\choose m}$ microstates $i(m)$ associated with $m$ are 
\be
\label{1'}
P_{i(m)} = P(m)/{50\choose m}.
\ee
If we substitute Eq.~(\ref{1'}) into Eq.~(\ref{1}) and do
the sum over
$i(m)$, we obtain
\be
\label{gibbs}
S = - \sum_{m=0}^{50} P(m)\ln P(m) +\sum_{m=0}^{50}P(m) \ln
{50\choose m} .
\ee
The second term on the
right-hand-side of Eq.~(\ref{gibbs}) arises, as was just shown, from
the fact that a macrostate having $m$ fleas on Anik has an
additional contribution to the entropy coming from the
equally likely microstates which make up the
macrostate. If the expression (\ref{0}) is
substituted into Eq.~(\ref{gibbs}), the result is the
previously mentioned $50\ln 2$.  We shall call this new entropy the
Gibbs entropy, because it is  analogous to the entropy in 
Gibbs's canonical ensemble. 

To associate a Gibbs entropy with the early, and consequently
non-equilibrium, part of the time development we have been
discussing, we need more information than the single time trace we
have been discussing. Because this entropy is a
property of a distribution, we need to assign probabilities to
every time step of the process, which means that we have to
contemplate an ensemble of subsystems and define probabilities in
terms of occurrences in the ensemble. One way to proceed would
be to create a very large number of traces such as the one in
Fig.~1, all of them starting with the same configuration.
Because the sequence of random numbers would be different in each
run, these traces would differ from one another. At
any given time, we could calculate a histogram like Fig.~3. 

Obtaining reliable distributions by this method would require a very
large number of runs. Fortunately, there is a much simpler way of
implementing the idea, which does not require a random number
generator. Let us calculate a distribution function $P(m)$, with
$m$ running from 0 to 50, which changes from step to step, and
reflects the random transfer of fleas from dog to dog. At the start
of the process we know with certainty that there are no fleas
on Anik. In the language of probability, the distribution at $t=0$
is
$P_0(0)=1, P_0(1)=P_0(2)= \cdots =P_0(50)=0$. (To indicate the
time we need another label, which we shall
write as a subscript.)  Now we argue that the
probability distribution at time $t$ determines the
probability distribution at time $t +1$. The assumption
that the fleas are being called at random implies that:
\be
\label{3}
P_{t+1}(m) = {{ m + 1}\over 50} P_t(m+1)~+~{{50 - (m-1)}\over 50} P_t(m-1).
\ee
Equation~(\ref{3}) can be understood by saying it in words. Anik can
have
$m$ fleas at time $t+1$ {\it either} because she had $m+1$ at time
$t$ and one jumped off, which has a probability proportional to
$m+1$, {\it or} because she had $m-1$ and one jumped on, which has a
probability proportional to the $50-(m-1)=51-m$ fleas that were on 
Burnside at time $t$.

We can write a program to develop the distribution corresponding 
to the initial certainty forward in time using Eq.~(\ref{3}).
However, there is one artificiality in this time evolution: at odd
(even) times only odd (even) numbers of fleas can be on Anik. This
artificiality can be remedied by averaging Eq.~(\ref{3}) over two
forward steps. The resulting evolution is shown in Fig.~5.

\begin{figure}
\centerline{\psfig{figure=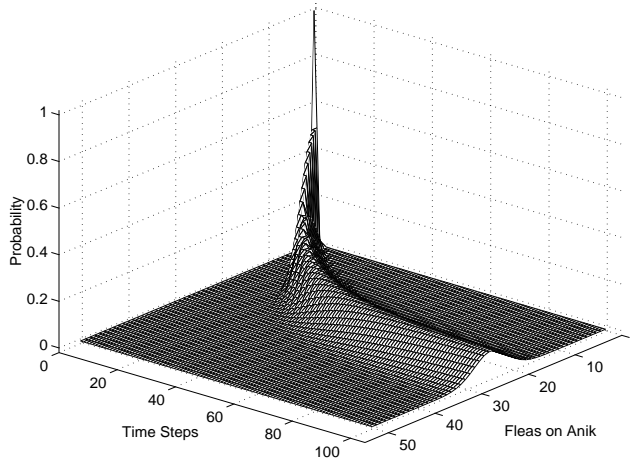,width=8.5cm} }
\caption{The probabilities on fleas on Anik with the passage of
time.}
\end{figure}

The three-dimensional plot in Fig.~5 is obtained by stacking
together the distributions at successive times. It very clearly
shows the initial
certainty evolving to a distribution --- which, not surprisingly,
can be shown to be the binomial in Eq.~(\ref{0}) --- with the outer
regions in the range of possibilities being extremely unlikely.
At each step we can calculate an entropy using
Eq.~(\ref{gibbs}). The result, shown in Fig.~6, shows that the
entropy rises steadily from zero to
$50\ln 2$.

By imagining that the system can be restarted at will, we have as in
Section~III insisted on the possibility of the external
imposition of an initial condition that is overwhelmingly unlikely
in equilibrium. Time-symmetric dynamics applied to such an initial
condition is overwhelmingly likely to evolve towards equilibrium.
\cite {PEIERLS} If it were possible to repeatedly arrange for
such a state to be a final condition, we could use a
backwards-in-time evolution equation relating the distribution
function at time $t-1$ to the distribution function at time $t$.
This equation would be identical to Eq.~(\ref{3}), except for the
replacement of $t+1$ by $t-1$:
\be
\label{33}
P_{t-1}(m) = {{ m + 1}\over 50} P_t(m+1)~+~{{51 - m}\over 50} P_t(m-1).
\ee

\begin{figure}
\centerline{\psfig{figure=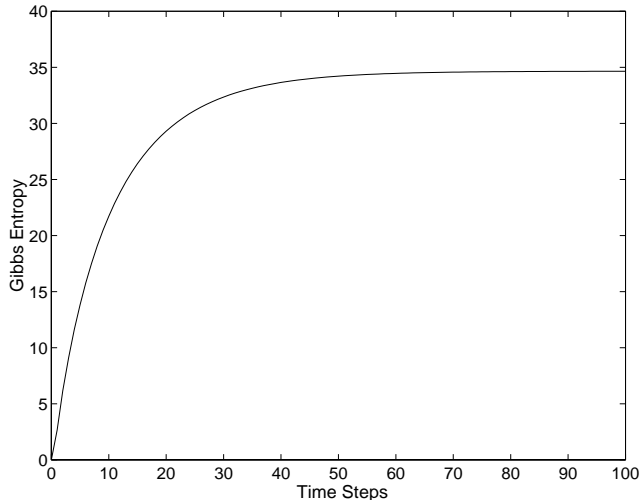,width=8.5cm} }
\caption{Entropy of Anik's fleas computed by averaging over many trials,
via the time-dependent probabilities generated via Eq.~(\ref{3})}
\end{figure}

Equation~(\ref{33}) would predict the opposite of what
is shown in Fig.~5 --- as one moved back in time through the history
of the  system, the entropy would increase
monotonically. But such  repeated occurrences of low entropy states
as fluctuations in equilibrium  are unimaginable. Such states do
not typically arise in this fashion, nor can we arrange for them to
do so. Time asymmetry in this context thus originates through our
use of Eq.~(\ref{3}) and rejection of Eq.~(\ref{33}).\cite{PEIERLS}

\section{ENERGY AND TEMPERATURE}

Up to now, the word temperature has only appeared at the
end of Section~III where it was argued that the standard Ehrenfest
model describes equilibration at high temperature. We have
confirmed this argument by showing that the entropy evolves to the
situation in which all microscopic configurations are equally
likely. It is, however, not difficult to introduce temperature
in this context. Suppose that Anik is cleaner than Burnside,
providing a less friendly environment for fleas. We may model this
environment by assuming an energy cost
$\epsilon$ to be paid by a flea jumping from Burnside to Anik. Let
the fleas be at an effective temperature $T$ (in energy units), and
define
$\Delta =
\epsilon/T$. We argue that 
Eq.~(\ref{3}) should be changed to
\be
\label{4}
 P_{t+1}(m) = {m+1\over 50}P_t(m+1)+ {50-m\over 50}\big[1
-e^{-\Delta}\big]P_t
(m)
+ {50-(m-1)\over 50}e^{-\Delta}P_t(m-1) .
\ee
For the new conditions, we expect that in
equilibrium any particular flea will spend more time on (dirty)
Burnside than on (clean) Anik. This expectation is implemented in
Eq.~(\ref{4}), which implies that when the number of one
of the fleas on Anik is called, it jumps to Burnside with
probability unity (term 1), but if one of the fleas on Burnside is
called, it either stays put with probability $1 - e^{-\Delta}$
(term 2), or jumps with probability $e^{-\Delta}$ (term 3).

Making the jump-probability from Burnside to Anik smaller than the
reverse process by the factor $e^{-\Delta}$ does in fact achieve
equilibrium in the steady state. In equilibrium, at temperature
$T$, the probability
$p$ of a flea being on Anik, and the  probability $1-p$ of one
being on Burnside should be given by the Gibbs distribution
\be
\label{5}
p={e^{-\Delta}\over 1 + e^{-\Delta}} \quad
\mbox{and} \quad 1 - p = \frac{1}{1 + e^{-\Delta}}~.
\ee
For 50 fleas the equilibrium probability distribution should be
the binomial corresponding to 50 tosses of an unfair coin
with outcome probabilities $p$ and $1-p$:
\be
\label{6}
P_{\rm eq}(m) = {50\choose m} p^m (1-p)^{50-m}.
\ee
It can be verified that
Eqs.~(\ref{5}) and (\ref{6}) are a stationary solution of
Eq.~(\ref{4}), namely that
substituting this form on the right reproduces it on
the left. In short, the effective temperature of the fleas
determines how many are willing
to put up with Anik's cleanliness. In the high temperature limit,
$\Delta\ll 1$, Eq.~(\ref {3}) is recovered. At very low
temperatures, $\Delta\gg 1$ and few fleas leave the snug comfort of
Burnside.

Several interesting and informative computations can be
performed with the evolution equation Eq.~(\ref{6}). We will focus
on one which we find particularly illuminating.  Shown in Fig.~7
are three entropy versus time traces, each at a different
temperature but with the same initial condition of all 50 fleas
on the ``clean''  (energetically unfavorable) dog Anik.  The curves have
been generated by computing the the Gibbs entropy (\ref{gibbs}) at
successive time steps.
We observe that at
low temperatures, the entropy of the  system does not increase
monotonically in time --- after a certain critical time, it actually
starts to {\it decrease}. Have we managed to violate the
second law?

\begin{figure}
\centerline{\psfig{figure=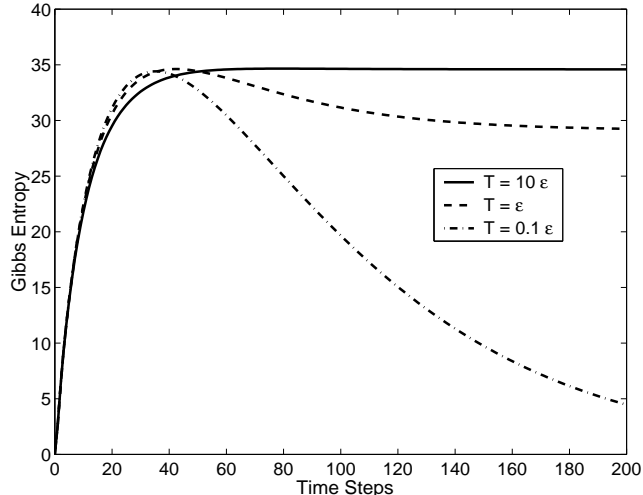,width=8.5cm} }
\caption{Evolution of the Gibbs entropy for various temperatures, starting
from a state where all fleas are on the ``clean'' dog Anik.}
\end{figure}

A little thought shows that there is no violation. The second law
requires only that the {\it total} entropy of the dog-flea system 
plus reservoir increase with time. The entropy of the reservoir is
insensitive to changes in the configuration of the fleas only at
temperatures much greater than the energy cost $\epsilon$. In
general, changes in entropy and energy of the reservoir at
temperature $T$ are related by 
\be
\label{8}
dS_{\rm res} = {dU_{\rm res}\over T}= - {dU\over T},
\ee
where the last equality is the result of conservation of 
energy, and $U$ is $\epsilon$ times the number of fleas on Anik.
(Note that the energy transfer is at at fixed $\epsilon$, which
implies that no work is done.\cite{VA,REIF}) Using Eq.~(\ref{8}),
an increase of the total entropy translates as usual into a
decrease of the Helmholtz
free energy $F$ of the flea subsystem, defined by 
$F = U - TS$,
where $T$ is the temperature of the reservoir and $S$ is the 
entropy defined in Eq.~(\ref{gibbs}). In Fig.~8, we plot the time
evolution of the free energy for the same initial conditions and
temperatures used in Fig.~7. We see that in all cases, the free
energy decreases monotonically  with time. 

Note that the initial condition for Figs.~7 and 8 (all fleas on
Anik) corresponds to having a negative temperature for the dog-flea
system. Consequently, reducing the internal energy (moving fleas
from Anik to Burnside) initially increases the entropy above its
equilibrium low temperature value.

\section{CONCLUSION}

We have demonstrated that an understanding of time in statistical mechanics 
can be obtained by
carefully examining the simple Ehrenfest dog-flea model. The model
has the virtues of offering qualitative insights and yielding easily to 
quantitative analysis. Our study has emphasized the manner in which
time reversal invariance is maintained in the model, and the role of
initial conditions in establishing a direction of time. We have
also shown that the model can be extended to finite
temperatures, where it may be used to explore interesting issues.

\begin{figure}
\centerline{\psfig{figure=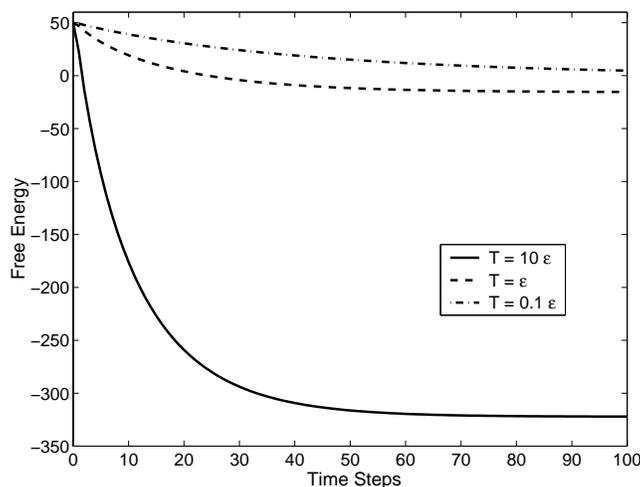,width=8.5cm} }
\caption{
Evolution of the free energy $F$ (in units of $\epsilon$) for various
temperatures, starting from
a state where all fleas are on the ``clean'' dog Anik.}
\end{figure}

Finally, we list some suggestions for further reading. Excellent elementary
discussions are to be found in Ref.~7.
The subject
is also treated in many textbooks accessible to advanced
undergraduates.
\cite{TEXTS} Whereas the topic is often underplayed in courses on
thermal physics, the opposite may be true in specialized
books. Several thought provoking articles as well as discussions of
the deep implications of the ideas presented here are to be found
in Ref.~\ref{bib:PHIL}.

\medskip
\noindent {\bf Added Note}:
Our colleague Ben Widom remarks that there are
`purists'---among whom he does not include himself---who think that the
Ehrenfest model is not a first-principles explanation of irreversibility
because there is a `stochastic element' in the model, which makes it `not
deterministic, as real dynamics is . . .'  To any such purists among our
readers, we point out that our implementation of the model uses
computer generated pseudo-random numbers which are completely
deterministic.
More generally, quite simple mechanical systems can generate pseudo-random
numbers, so that the Ehrenfest heat reservoir can be thought of in a
completely mechanical way. (See Chapter 11 of Ref. 3 for an elementary
introduction to deterministic chaos.)
 
\bigskip
\centerline {\bf ACKNOWLEDGEMENTS}
\medskip

This work has been partially supported by the National Science
Foundation under grant DMR-9805613.


\begin{references}
\bibitem{FTNT} But not perfect. How reasonable the disagreement is,
given the number of trials, is a question we desist from asking,
because ultimately that would include among other things a test of
the random number generator.
\end{references}
\end{document}